\documentclass[prl,twocolumn,showpacs,amsmath,amssymb]{revtex4}
\usepackage{graphicx}
\usepackage{color}

\begin{document}
\title{Switching off the magnetic exchange coupling by quantum resonances}
\author{Ching-Hao Chang}\email{cutygo@gmail.com} 
\author{Tzay-Ming Hong}\email{ming@phys.nthu.edu.tw}
\affiliation{Department of Physics, National Tsing Hua University, Hsinchu, Taiwan 300, Republic of China}
\date{\today}

\begin{abstract}
We clarify the role of quantum-well states in magnetic trilayer systems from majority carrier in the ferromagnetic and all carriers in the antiferromagnetic configurations. In addition to numerical and analytic calculations, heuristic pictures are provided to explain effects of a capping layer and side-layer modulation in recent experiments. This immediately offers answers to two unexplained subtle findings in experiments and band-structure calculations, individually. Furthermore, it allows a more flexible tuning of or even turning off the interlayer exchange coupling.
\end{abstract}

\maketitle
It has been observed\cite{gmr1,gmr2} more than twenty years ago that the interlayer exchange coupling (IEC) of ferromagnetic layers in giant magnetoresistance (GMR) system
oscillates between ferromagnetism and anti-ferromagnetism as a function of spacer width. This feature is ascribed to the Ruderman-Kittel-Kasuya-Yosida (RKKY) oscillation \cite{rkky,md}
and the formation of quantum well states (QWS) at Fermi surface \cite{qmstate1,qmstate2,qmstate3}. These theories, based on the ideal system with infinite system widths,
are successful at predicting the IEC periods.
However, the theoretic values for their amplitudes always overestimate.
Although the inevitable roughness on the interfaces was considered as the key factor to this disagreement in most cases\cite{rough},
the treatment of approximating the layer thickness as being semi-infinite should also play an important role\cite{brunosl,weight}. 
The \emph{ab initio} calculation\cite{weight} in Co/Cu/Co (100) and (010) has revealed that the coupling strength is very sensitive to the variation of side-layer (SL) thickness,
which redistributes weightings among different RKKY modes. 
The physical origin in (100) case was ascribed to the discrepancies between the Co density of state in the bulk and that in the thin layer, but in (010) the density of states is roughly the same and so it was put forward as a subtle challenge.

In addition, the experimentalists\cite{capping,capping2,capping3} found that the coupling strength displayed an oscillatory feature as the SL thickness was varied.
It was discovered recently that, with an insulating capping layer and by varying SL thickness, the scattering properties in the SL edge became tunable\cite{capping}. In contrast to the usage of a metallic cap, the oscillation amplitude got stronger and a novel RKKY period was deduced\cite{capping}.
The quantum resonance, which is altered by the scattering paths reflected between the bottom of capping layer and the top of spacer, would be crucial
to answer (1) the origin of disagreements for the noble-metal spacer\cite{md} and (2) the creation of new IEC period observed in recent experiments\cite{capping}.
Another motivation for studying the finite-size effect for IEC is that
it may enable us to either enhance or shut off entirely the IEC by proper nano-configurations.
A known example of such applications is the diluted magnetic semiconductors (DMS), where a room-temperature ferromagnetic phase has been obtained
by replacing the magnetic impurities with the self-organized nanocolumns\cite{cluster}.
The quantum resonance in these nanocolumns, which is realized at properly chosen sizes, enhances the RKKY coupling between clusters and achieves a higher transition temperature for the DMS\cite{mydms}.

To realize how the boundaries with the capping layer and substrate for finite ferromagnetic layers affect the IEC,
we start with the GMR consisting of a noble-metal spacer, in
which the minority carrier are confined in the spacer and their quantum resonance near the Fermi surface will determine the position of peaks in the IEC oscillation\cite{qmstate2,noble-metal}.
The Fe/Ag/Fe (001) trilayer is calculated within the single-electron picture with different energy barriers in respective layers\cite{smith}.
Besides, we assume that both the substrate and the capping layer play the role of infinite potential barriers to simulate the recent experimental setup\cite{capping}.
The exchange coupling is defined as the relative energy difference $J=(\Delta E_{AFM}-\Delta E_{FM})/A$ with $A$ being the interface area.
The $\Delta E_{FM}/\Delta E_{AFM}$ denotes the system energy of normal order for the FM/AFM configuration. For the case of FM, this energy is calculated by
\begin{align}
\notag\Delta E_{FM}&=A E_{F}\sum_{\sigma=\uparrow, \downarrow}\sum_{n=0}^{\infty}\int d^{2}k_{\|}/4\pi^{2}\\
&\frac{E(\sigma,n)/E_{F}+k^{2}_{\|}/k^{2}_{F}}{exp\Big[\big(E(\sigma,n)/E_{F}+k^{2}_{\|}/k^{2}_{F}-1\big)\frac{E_{F}}{k_{B}T}\Big]+1},
\label{eq:energy}
\end{align}
where $E(\sigma,n)$ is the energy of {\it n}-th bound state for carriers with spin orientation $\sigma$ in the confined trilayer, $k_{\|}$ is the transverse wave vector, and $T$ is the temperature.
Note that $E_{F}/k_{F}$ is assigned to denote the Fermi energy/momentum for the spacer and $q_{F}$ is the Fermi momentum for the majority carrier in SL from now on.

\begin{figure}[h!]
\includegraphics[width=0.4\textwidth]{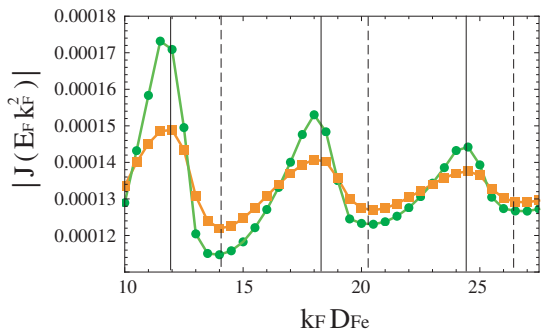}
\caption{(color online) Coupling energies for Fe/Ag/Fe (001) trilayer with a fixing $D=5/k_{F}$ are estimated  for different SL widths. 
Both SL are of equal finite width for the circle points. In contrast, the square points represents some experiment conditions where one of the SL is semi-infinite, which we mimic by setting $D=200/k_{F}$. The solid/dashed vertical lines are the theoretic prediction where the peaks/bottom values will occur, on which more discussions will follow in Fig.\ref{fig-state}.}
\label{fig-jfe}
\end{figure}

\begin{figure}
\includegraphics[width=0.4\textwidth]{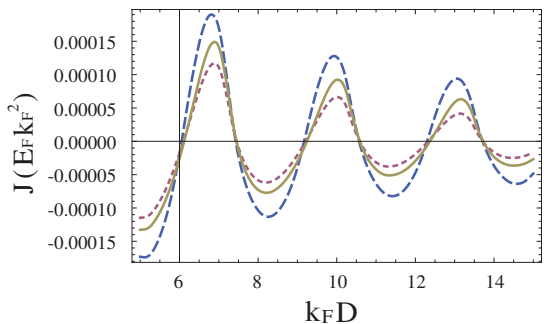}
\caption{(color online) Coupling energies plotted as a function of  $D$ for  Fe/Ag/Fe (001). The solid line denotes the case where both SL are set at $k_{F}D_{{\rm Fe}}=200$ to mimic the semi-infinite system.
The dashed/dotted lines are for finite systems where $k_{F}D_{{\rm Fe}}=11.5/14$ corresponds to the first peak/bottom of the solid line in Fig.\ref{fig-jfe}. }
\label{fig-comparison}
\end{figure}

To describe Fe/Ag/Fe (001) system in interface zone center at $k_{B}T=10^{-4} ev$, Eq.(\ref{eq:energy}) and the coupling strength $J$ can be estimated by setting $E_{F}=4ev$, $1/k_{F}=4\AA$  and the potential barrier for majority/minority carrier in SL is $5/3 ev$\cite{smith,md2}.
In Fig.\ref{fig-jfe}, the spacer width $D$ is fixed while $J$ is plotted as a function of SL thickness, $\rm D_{{\rm Fe}}$.
Consistent with the experimental findings, the oscillation period is dominated by and equals half of the Fermi wavelength of majority carrier in the ferromagnetic side layer.
Similar to the conventional RKKY oscillation,  $J$ also displays the power-law decay and approaches the result of a semi-infinite system as  ${\rm D_{{\rm Fe}}}\rightarrow \infty$.
In Fig.\ref{fig-comparison}, we study  the effect of $D$ on  $J$ while the SL width is fixed at the values which correspond to one of the peaks and bottoms respectively in Fig.\ref{fig-jfe}.
Although the extreme values of $J$ occur at roughly the same spacer widths, which is consistant with experiments\cite{qmstate3,ZQ},
their amplitudes are sensitive to the specific choice of SL thickness, with a possible enhancement of doubling the semi-infinite value at $k_{F}D>14$.
Overall, the thicker the $D$, the larger  the enhancement. 

\begin{figure}[h!]
\includegraphics[width=0.45\textwidth]{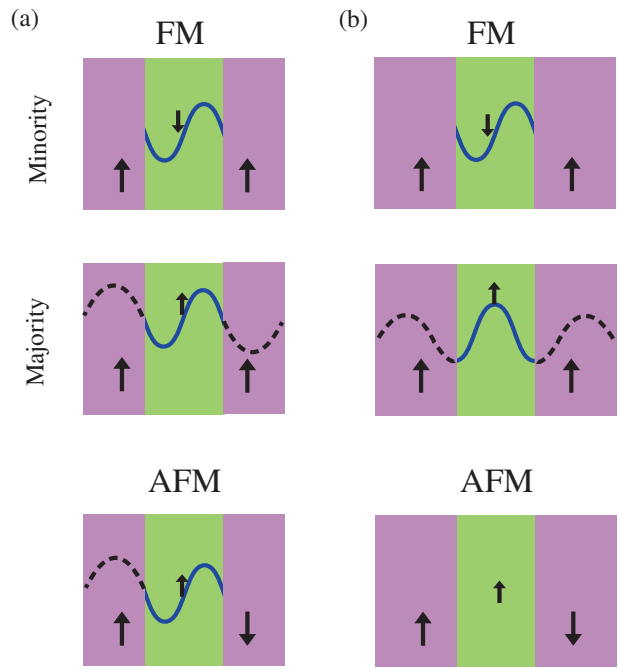}
\caption{(color online) Schematic diagrams that exemplify (a) the coherent case, which characterizes the simultaneous appearance of quantum resonances for carriers in FM and AFM configurations and (b) the incoherent case by adding a quarter Fermi wavelength to both SL thickness. All carriers in AFM no longer sustain QWS due to the violation of boundary conditions. The big/small arrow denotes the moment/spin orientation  of SL/carrier. }
\label{fig-diagram}
\end{figure}

In the semi-infinite case which contains noble-metal spacer, Oretega {\it et al} \cite{qmstate1,qmstate2} noticed that the minority state or equivalently QWS could be generated periodically as a function of spacer width, while  the unbound majority state is nearly unaffected.
The extreme values of $J$ appear as the minority carrier form QWS.
However, all realistic systems are of finite size; esp. when in nanoscale the confinement and formation of QWS in FM/AFM configurations come from not just the minority but all carriers. 
 One interesting consequence is that the amplitude of $J$ becomes tunable by adjusting the SL thickness, as shown in Fig.\ref{fig-comparison}. Although being unable to derive this, Ref.\cite{qmstate1,qmstate2} correctly predicted the special spacer widths that could render the extreme values. The schematic diagrams in Fig.\ref{fig-diagram} provide heuristic pictures for understanding how SL thickness alters the relative positions of quantum resonances for different carriers and makes such a manipulation possible. 
The energy barrier height in the side layer is set to be zero/infinity for the majority/minority carrier for convenience without loss of generality.

\begin{figure}[h!]
\includegraphics[width=0.4\textwidth]{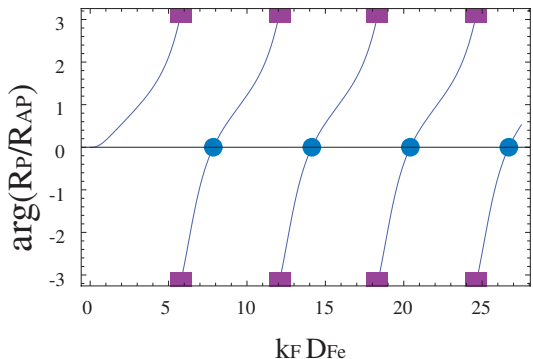}
\caption{(color online) The argument of $R_{P}/R_{AP}$ is plotted as a function of SL thickness. Since $|R_{P}|=|R_{AP}|=1$, the zero arguments (in circles) denote the case of $R_{P}=R_{AP}$ which allows coherent resonances. While $\pm \pi$ arguments (squares) are for $R_{P}=-R_{AP}$ that induce incoherent resonances.}
\label{fig-state}
\end{figure}

The SL thickness in Fig.\ref{fig-diagram}(a) allows constructive interferences for the majority carrier in FM (see middle plot) and for both carriers in AFM (bottom plot) configurations.
Since these resonances are absent in the semi-infinite case, they enhance the system energy of both configurations by an amount that is roughly double in AFM than in FM. When taking their difference, $J$ turns out to be weaker. In Fig.\ref{fig-diagram}(b), we add a quarter Fermi wavelength to both SL thickness. The resonance is ruined in AFM configuration, but is kept alive in FM. This results in an enhanced $J$ as compared to the semi-infinite case.
Irrespective to whether the capping layer and substrate are metallic or insulating, the effective reflection coefficient at both edges of the spacer can be calculated after the inclusion of additional scattering paths that are reflected by them
\begin{align}
R_{P}(r_{c},D_{{\rm Fe}})=\frac{r_{P}+r_{c}e^{2iq_{F}D_{{\rm Fe}}}}{1+r_{P}r_{c}e^{2iq_{F}D_{{\rm Fe}}}},
\label{eq-effectiveR}
\end{align}
where $r_{P}$ and $r_{c}$ denote the reflection coefficient for the majority carrier reflected by SL and the capping layer, respectively. 
The  coefficient $R_{AP}$ for the minority carrier obey a similar equation as Eq.(\ref{eq-effectiveR}) except the wavevector $q_{F}$ in SL becomes complex.
The condition for the formation of QWS is $R^{2}_{P/AP}e^{2ik_{F}D}=1$ for majority/minority carrier in FM, in contrast to $R_{P}R_{AP}e^{2ik_{F}D}=1$ for both carriers in AFM configuration.	
These three relations allow us to deduce the necessary condition for the coherent case as $R_{P}=R_{AP}$ as opposed to $R_{P}=-R_{AP}$ for the incoherent one.
The connection between the hugely tunable $J$ and our heuristic picture of an incoherent/coherent appearance of QWS
is confirmed by comparing Fig.\ref{fig-state} with the SL thicknesses that correspond to an enhancement/reduction of $J$ in Fig.\ref{fig-jfe}.

\begin{figure}
\includegraphics[width=0.37\textwidth]{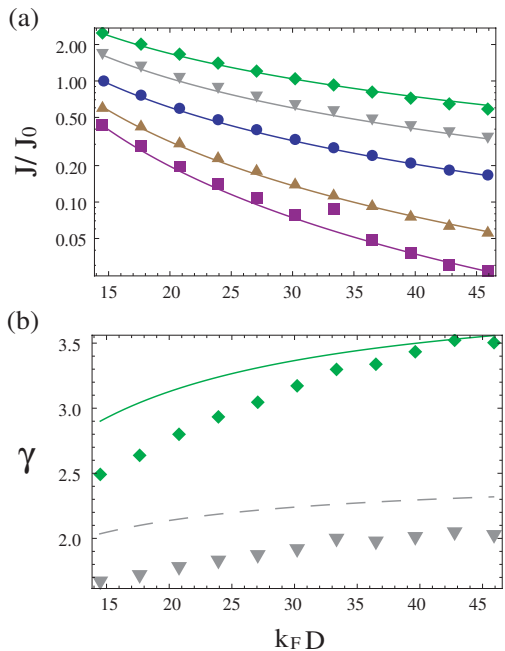}
\caption{(color online) Panel (a): The peak values of coupling strength for Fe/Ag/Fe (001) with different arrangements of SL thickness: semi-infinite (circle), of the same thickness $k_{F}D_{{\rm Fe}}=5.79/7.85$ (diamond/square), and with one obeying $k_{F}D_{{\rm Fe}}=5.79/7.85$  while the other is semi-infinite (inverse triangle/triangle). The unit $J_{0}$ is the coupling strength for the semi-infinite case with $k_{F}D=14.5$. All five lines are fitted by a power-law decay with exponent $-1.18, -1.38, -1.56, -2.06$, and $-2.41$ from up to bottom. 
Panel (b): Same plot for the enhancement ratio with the same thickness $k_{F}D_{{\rm Fe}}=5.79$ for both SL (diamond) and with one SL set to be semi-infinite (inverse triangle). The solid and dashed lines are plotted from Eqs.(\ref{eq-gamma}) and (\ref{eq-gamma2}). }
\label{fig-asym}
\end{figure}

In order to study the enhancement in more detail, we start from the asymptotic form of IEC within the free electron model and assume that SL are made by the same material\cite{rkky}:
\begin{align}
\lim_{k_{F}D\gg 1} J \approx\frac{-E_{F}k_{F}^{2}}{8\pi^{2}(k_{F}D)^{2}}{\rm Im}\Big[\big(r_{P}-r_{AP}\big)^{2}e^{2ik_{F}D}\Big].
\label{eq:ja}
\end{align}
By expanding  $\big(r_{P}-r_{AP}\big)^{2}$, the quantum interferences for $r_{AP}^2$, $r_{P}^2$, and -2$r_{AP}r_{P}$ terms are each characterized in top, middle, and bottom plots in Fig.\ref{fig-diagram}.
Quantitatively,  $|r_{AP}|\approx 1$ is much bigger than $|r_{P}|$ in the semi-infinite case, as opposed to being exactly equal when finite and sustaining QWS. 
This implies that the maximum enhancement from the incoherent resonance $R_{P}=-R_{AP}$ could reach as high as four times in the semi-infinite case.
This prediction is verified in Fig.\ref{fig-asym}(a).
In Fig.\ref{fig-comparison}, it was observed that the enhancement ratio got bigger as $D$ increases. This can also be explained by examining the denominator of Eq.(\ref{eq:ja}). For finite SL, the contribution from the majority carrier in FM needs to be modified to  $R_{P}^2/(k_{F}D+2q_{F}D_{{\rm Fe}})^{2}$,  while that from the minority remains the same. In contrast, the contribution from both carriers in AFM configuration is changed to to $-R_{AP}R_{P}/(k_{F}D+q_{F}D_{{\rm Fe}})^{2}$. 

Let's now concentrate on the incoherent case. When $D\gg 2(q_F/k_F)D_{{\rm Fe}}$, the contribution from all carriers becomes roughly $R_{AP}^2/(k_{F}D)^{2}$ and equals to that of the minority. On the other hand, it is much smaller when $D\ll 2(q_F/k_F)D_{{\rm Fe}}$. So one should adope a thick spacer in order to obtain a large enhancement ratio, $J(D)/J(D\ll 2(q_F/k_F)D_{{\rm Fe}})$, which depends on
\begin{align}
\gamma \approx 1+\Big(\frac{k_{F}D}{k_{F}D+2q_{F}D_{{\rm Fe}}}\Big)^{2}+2 \Big(\frac{k_{F}D}{k_{F}D+q_{F}D_{{\rm Fe}}}\Big)^{2}.
\label{eq-gamma}
\end{align}
In the above case, all carriers are totally reflected at both edges and satisfy $R_{P}=-R_{AP}$. 
But when the substrate is semi-infinite or the top of GMR is either connected to a contact or capped by a metal layer, this side of SL should be treated as being semi-infinite. It will adopt the coefficients $r_{P}$ and $r_{AP}$ for the semi-infinite case, and the path factor for additional resonances all equals to $k_{F}D+q_{F}D_{{\rm Fe}}$.
As a result, Eq.(\ref{eq-gamma}) is modified to
\begin{align}
\gamma'\approx 1+\big|r_{ap}+2r_{p}\big| \Big(\frac{k_{F}D}{k_{F}D+q_{F}D_{{\rm Fe}}}\Big)^{2}.
\label{eq-gamma2}
\end{align}
It is obvious that Eq.(\ref{eq-gamma2}) is always less than Eq.(\ref{eq-gamma}) when $D$ is large, which explains why the amplitude of the square line in Fig.\ref{fig-jfe} is smaller than that of the circle one.
Both Eqs.(\ref{eq-gamma}) and (\ref{eq-gamma2}) are consistent quantitatively with the numerical results in Fig.\ref{fig-asym}(b). 
The nearly constant shift between Eq.(\ref{eq-gamma2}) and our calculation is due to the fact that the SL thickness we chose only generates approximate incoherent resonance in the system with one semi-infinite SL.

A final implication is that we may have solved the puzzle put forward by Halley {\it et al.} in Ref.\cite{capping}. By capping Fe/Cr/Fe, they managed to observe a novel period for $J(D)$, long predicted and sought after by the band-structure calculations\cite{md2}. As was argued by the authors of Ref.\cite{capping}, we provide a concrete physical process to explain how this additional confinement can lead to revival or suppression of the contribution to IEC from a particular channel via  modifications of the reflection coefficient at Fe/Cr. 
Equations (\ref{eq-gamma}) and (\ref{eq-gamma2}) have concentrated on the enhancement effect. Let us now derive in more detail how the suppression comes about. 
Mainly, one needs to create the coherent conditions mentioned in Fig.\ref{fig-diagram} (a).
Coupling strength of  the square line, which represents the coherent case, has the highest decaying rate than the other four lines in Fig.\ref{fig-asym}(a).
Its magnitude can be about one and two orders smaller than the circle (for the semi-infinite case) and the diamond lines (incoherent case) at the largest $D$ value in the figure, respectively. 
This weak coupling that results from this nonconventional RKKY oscillation, which is generated by $R_{P}=R_{AP}$, can be estimated by converting the plus sign for the third term in Eq.(\ref{eq-gamma}) to minus:
\begin{align}
J(D\gg 2\frac{q_F}{k_F}D_{{\rm Fe}})
\approx -\frac{E_{F}D^{2}_{Fe}}{4\pi^{2}}\frac{\sin (2k_{F}D+\phi)}{D^{4}}.
\label{eq:ja2}
\end{align}
The reason why this decaying power is higher than those obtained numerically in Fig.\ref{fig-asym}(a) is that the asymptotic form in Eq.(\ref{eq:ja}), on which Eq.(\ref{eq:ja2}) is based,  is only an approximation and the higher-order terms are not rigorously negligible for trilayers with noble-metal spacer\cite{non-asym}.  

Although there are many factors that may affect the RKKY oscillation in the real Fe/Cr/Fe sample,
our calculation confirms that the resonances in  both SL have the capability to drastically change the IEC strength. 
Another noticeable property is that GMR has multi-period RKKY oscillations, which is common in most experiments, and these
modes can be separately tuned by varying the SL thickness; i.e., changing their resonant conditions.
Their relative weightings are also affected by the SL and capping layer thickness.
We are confident that the predictions and explanations above for Fe/Ag/Fe can be equally applied to Co/Cu/Co. Nevertheless, a quantitatively match will require a more detailed \emph{ab initio} calculation than the pioneering work by Nordstr\"{o}m {\it et al.}\cite{weight}. 
For instance, they calculated the density of states for a single Co SL and showed that different thickness could lead to redistribution of RKKY modes. However, they could not locate the origin of this dependence in the (010) crystal direction since the density of states they obtained was not sensitive to varying SL thickness. Based on our predictions, we propose to redo the calculations for the whole trilayer; namely, include the spacer and both SL.

In conclusions, we find that the strength of IEC for GMR with a capping layer can be modulated by as much as two orders of magnitude through careful arrangement of SL thickness.
This is made possible by creating coherent (to suppress IEC) or incoherent (to enhance) QWS in FM and AFM configurations. 
This immediately offers answers to two outstanding puzzles in experiments\cite{capping} and band-structure calculations\cite{weight} individually. We provided heuristic physical pictures and both numerical and analytic calculations to support our conclusions. Moreover, the discrepancy between theoretic predictions for IEC strength of trilayers and experimental findings, which was ascribed to the interface roughness\cite{md}, also receives an alternative explanation. Our mechanism carries another potential merit at enabling experimentalists to eliminate the intergranular exchange coupling in the hard disc drives\cite{hd} and other future spintronic devices. 

\begin{acknowledgments}
We thank K. Lenz, M. K\"{o}rner, J. McCord, C. R. Chang, M. T. Lin, C. M. Wei, C. H. Lai and C. C. Chi for valuable comments and discussions. The authors acknowledge support by the NSC in Taiwan and DAAD in Germany. 
\end{acknowledgments}

\end{document}